\documentclass[showpacs,preprintnumbers]{revtex4}
\usepackage{graphicx}
\usepackage{dcolumn}
\usepackage{bm}

\begin{document}

\preprint{APS/123-QED}
\title{A scheme for generating entangled cluster state of atomic ensembles%
}
\author{Ya-Fei Yu}
\email{yfyuks@yahoo.com.cn}
\affiliation{Laboratory of Photonic Information Technology, South China Normal
University, Guangzhou 510006, PR China}
\author{Feng Mei}
\affiliation{Laboratory of Photonic Information Technology, South China Normal
University, Guangzhou 510006, PR China}
\author{Zhi-Ming Zhang}
\email{zmzhang@scnu.edu.cn}
\affiliation{Laboratory of Photonic Information Technology, South China Normal
University, Guangzhou 510006, PR China}
\author{Shi-Liang Zhu}
\affiliation{Institute for Condensed Matter Physics, South China Normal University,
Guangzhou 510006, PR China}
\date{Nov.11,2008}

\begin{abstract}
It was shown in Ref.[Phys.Rev.A 77,045802(2008)] that the dynamics
of a control atom and an atomic sample interacting dispersively with
a cavity can be discribed by the Jaynes-Cummings model and the
collective mode of the atomic sample can be analogous with a bosonic
mode. Here, by analogizing the behaviour of the atomic sample with
the one of the cavity, we propose a scheme to generate cluster
states of atomic ensembles by Cavity QED.
\end{abstract}

\maketitle

The phenomena of superposition and entanglement in quantum system
can provide enhancements on the ability of processing information
over what is possible or impossible classically, such as quantum
computation \cite{xue1}, quantum teleportation \cite{xue2},
superdense coding \cite{xue3}, and quantum cryptography \cite{xue4}.
Generation of entanglement between atoms, photons or combinations is
at the heart of quantum information science. The technology of
generation and manipulation of several partite entangled states has
been realized in many systems \cite{xue5}. There still has been much
interest in using quantum resource to get more and more subsystems
entangled \cite{xue6} for more applications. It has been shown that
there are several inequivalent classes of entangled states
\cite{zou1}. Recently, Briegel and Raussendorf introduced a class of
multi-qubit entangled states, i.e. the so-called cluster states
\cite{zou2}, which have some interesting features. They have a high
persistence of entanglement, and can be regarded as a resource for
the GHZ states \cite{zou2}, but they are more immune to decoherence
than the GHZ states \cite{zou3}. The cluster states have been shown
to violate a new Bell inequality which is not violated by the GHZ
state \cite{zou4}. And more importantly, the cluster state has been
reported to constitute a universal resource for one-way quantum
computation with proceeding only by local measurements \cite{zou5},
thereby eliminating the troublesome requirement for dynamically
controlled two-qubit operations. Much effort has been devoted to
generating the cluster state of single-particle systems
\cite{cluster1}. Atomic ensembles have been proposed to be a
promising candidate for implementations of quantum information
processing by many recently discovered schemes \cite{ae1}. The
schemes based on atomic ensembles have some special advantages
compared with the schemes on the control of single particles, such
as easier laser manipulation, collectively enhanced coupling
\cite{ae2}. In the last few years, much attention has been payed to
generate substantial spin squeezing \cite{ae3}
and continuous variable entanglement \cite{ae4}, store quantum light \cite%
{ae5}, realize scalable long-distance quantum communication \cite{ae6} and
prepare many-party entanglement \cite{ae7}. However, there is no any scheme
for generation of the cluster state in this system.

Recently a result has been shown in Ref.\cite{zheng} that, putting a single
control atom driven by an auxiliary classical field and an atomic sample
into a nonresonant cavity, under certain conditions the coherent energy
exchange between the control atom and the atomic sample can be described by
an effective Jaynes-Cummings model, where the collective ensemble atomic
spin is treated as a bosonic mode. This dynamics provide us an alternative
method to entangle in a single step a control atom with a mesoscopic number
of atoms. Based on the dynamics, we describe a scheme of preparing the
cluster state between many atomic ensembles in this paper. The scheme
involves the microwave cavity quantum electrodynamics (CQED) with a Rydberg
control atom crossing cascaded microwave cavities as proposed in Ref.\cite%
{zou}, each containing a Rydberg atomic ensemble. The schematic of the
scheme is given in Fig. 1. The basic element of this scheme is a Rydberg
atom and some ensembles of $N$ identical Rydberg atoms. The relevant levels
of the Rydberg atom are labelled by $\left\vert f\right\rangle ,$ $%
\left\vert g\right\rangle ,$ $\left\vert e\right\rangle $ (see Fig. 2). The $%
\left\vert g\right\rangle \leftrightarrow \left\vert e\right\rangle $ and $%
\left\vert g\right\rangle \leftrightarrow \left\vert f\right\rangle $
transitions are at the frequency $\omega _{0}$ and $\omega _{1}$,
respectively.

\begin{figure}[tbp]
\includegraphics[height=2.5cm]{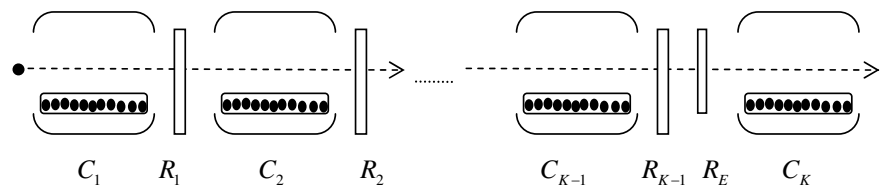}
\caption{Schematic setup to generate the cluster states of $K$ atomic
samples. $C_{i}(i=1\sim K)$ denotes the cavity $i$ including an atomic
sample and $R_{i}(i=1\sim K-1,E)$ the Ramsey zone.}
\end{figure}

\begin{figure}[tbp]
\includegraphics[width=4cm]{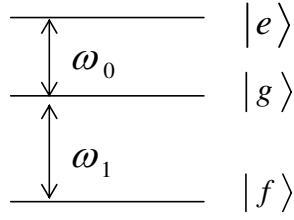}
\caption{The relevant level structure for the Rydberg atom, and the
transitions $\left\vert g\right\rangle \leftrightarrow \left\vert
e\right\rangle $ and $\left\vert g\right\rangle \leftrightarrow \left\vert
f\right\rangle $ are resonant at the frequency $\protect\omega _{0}$ and $%
\protect\omega _{1}$, respectively.}
\end{figure}

The first step of the scheme is to feed the atomic sample and the control
atom into the first microwave cavity. the control atom is illuminated by an
auxiliary classical field. The cavity frequency is appropriately chosen in a
way that only the transition $\left\vert g\right\rangle \leftrightarrow
\left\vert e\right\rangle $ is affected by the cavity, the state $\left\vert
f\right\rangle $ is not affected during the atom-cavity interaction. So in
the first cavity the Hamiltonian for the whole system is (assuming $\hbar =1$%
) \cite{zheng}%
\begin{equation}
H=H_{0}+H_{i}\text{,}
\end{equation}%
where%
\begin{equation}
H_{0}=\omega _{c}a^{+}a+\omega _{0}(S_{z,c}+\sum\limits_{j=1}^{N}S_{z,j})%
\text{,}
\end{equation}%
\begin{equation}
H_{i}=\Omega (S_{c}^{-}e^{i\omega _{L}t}+S_{c}^{+}e^{-i\omega
_{L}t})+g(a^{+}S_{c}^{-}+aS_{c}^{+})+g\sum%
\limits_{j=1}^{N}(a^{+}S_{j}^{-}+aS_{j}^{+})\text{,}
\end{equation}
$a^{+}$ and $a$ are the creation and annihilation operators for the cavity
mode, $S_{z}=\frac{1}{2}(\left\vert e\right\rangle \left\langle e\right\vert
-\left\vert g\right\rangle \left\langle g\right\vert ), S^{+}=\left\vert
e\right\rangle \left\langle g\right\vert , S^{-}=\left\vert g\right\rangle
\left\langle e\right\vert $ are the inversion, rising, and lowering
operators for the atom, the subscripts $c$ and $j$ denote the control atom
and the $j$th atom in the sample, respectively. $\omega _{L}$ and $\Omega $
are the oscillation frequency and the Rabi frequency of the classical field,
respectively. $\omega _{c}$ is the cavity frequency, $g$ is the atom-cavity
coupling strength. Under the detuning condition $\delta _{c}=\omega
_{0}-\omega _{c}\gg g\sqrt{N(\overline{n}+1)}$, with $\overline{n}$ being
the mean photon number of the cavity field, there is no energy exchange
between the atomic system and the cavity. The dispersive atom-cavity
interaction leads to photon-number-dependent Stark shifts and dipole
couplings for the atomic system. And when $\delta _{L}=\omega _{0}-\omega
_{L}\gg \Omega $, the classical field only induces a Stark shift \cite{zheng}%
. If the cavity is initially in the vacuum state, it will remain in the
state. So in the following the state of the cavity is omitted. The effective
Hamiltonian of the whole system reduces to \cite{zheng}%
\begin{eqnarray}
H_{e} &=&\lambda _{L}(\left\vert e_{c}\right\rangle \left\langle
e_{c}\right\vert -\left\vert g_{c}\right\rangle \left\langle
g_{c}\right\vert )+\lambda _{c}(\left\vert e_{c}\right\rangle \left\langle
e_{c}\right\vert +\sum\limits_{j=1}^{N}\left\vert e_{j}\right\rangle
\left\langle e_{j}\right\vert ) \\
&&+\lambda
_{c}(\sum\limits_{j=1}^{N}(S_{c}^{+}S_{j}^{-}+S_{c}^{-}S_{j}^{+})+\sum%
\limits_{j,k=1,j\neq k}^{N}S_{k}^{+}S_{j}^{-})\text{,}  \nonumber
\end{eqnarray}%
where $\lambda _{L}=\frac{\Omega ^{2}}{\delta _{L}}$ and $\lambda _{c}=\frac{%
g^{2}}{\delta _{c}}$. Introduce the operators $b=\frac{1}{\sqrt{N}}%
\sum\limits_{j=1}^{N}S_{j}^{-}, b^{+}=\frac{1}{\sqrt{N}}\sum%
\limits_{j=1}^{N}S_{j}^{+}, n_{b}=\sum\limits_{j=1}^{N}\left\vert
e_{j}\right\rangle \left\langle e_{j}\right\vert .$ By discarding a constant
energy $\frac{\lambda _{c}}{2}$, the Hamiltonian of the system can be
rewritten as \cite{zheng}%
\begin{equation}
H_{e}=(2\lambda _{L}+\lambda _{c})S_{z,c}+N\lambda _{c}b^{+}b+\sqrt{N}%
\lambda _{c}(S_{c}^{+}b+S_{c}^{-}b^{+}).
\end{equation}%
From the definition it is obvious that $[b,b^{+}]=1-\frac{2}{N}n_{b},$ $%
[n_{b},b^{+}]=b^{+},[n_{b},b]=-b$. Therefore, if $N\gg 1,\overline{n}_{b}$,
with $\overline{n}_{b}$\ being the average excitation number of the atomic
sample, $b$ and $b^{+}$ can be regarded as the bosonic operators and the
atomic sample can be taken for a bosonic system. The Hamiltonian $H_{e}$ in
Eq.(5), resembling the Jaynes-Cummings Hamiltonian, describes the
oscillatory exchange of an excitation between the control atom and the
collective atomic mode.

Suppose that $2\lambda _{L}=(N-1)\lambda _{c}$, the above Hamiltonian gives
rise to the evolution of the state of the atomic system as
\begin{eqnarray}
\left\vert e_{c}\right\rangle \left\vert n\right\rangle &\longrightarrow
&\cos (\sqrt{\left( n+1\right) N}\lambda _{c}t)\left\vert e_{c}\right\rangle
\left\vert n\right\rangle -i\sin (\sqrt{\left( n+1\right) N}\lambda
_{c}t)\left\vert g_{c}\right\rangle \left\vert n+1\right\rangle \text{,} \\
\left\vert g_{c}\right\rangle \left\vert n+1\right\rangle &\longrightarrow
&\cos (\sqrt{\left( n+1\right) N}\lambda _{c}t)\left\vert g_{c}\right\rangle
\left\vert n+1\right\rangle -i\sin (\sqrt{\left( n+1\right) N}\lambda
_{c}t)\left\vert e_{c}\right\rangle \left\vert n\right\rangle \text{,}
\nonumber
\end{eqnarray}%
with $\left\vert n\right\rangle $ denoting the Fock-like state for the
collective atomic mode.

Assuming the control atom is initially prepared in the superposition state $%
\frac{\left\vert f_{c}\right\rangle +\left\vert e_{c}\right\rangle }{\sqrt{2}%
},$ the state of the collective atomic mode is $\left\vert 0\right\rangle $,
i.e., all the atoms in the sample stay in the state $\left\vert
g_{j}\right\rangle .$ That is, the initial state of the atomic system is $%
\frac{\left\vert f_{c}\right\rangle +\left\vert e_{c}\right\rangle }{\sqrt{2}%
}\left\vert 0\right\rangle $. After the control atom passed through the
first cavity with an interaction time $t=\frac{\pi }{2\sqrt{N}\lambda _{c}}$%
, the whole state of the control atom and the atomic sample becomes%
\begin{equation}
\frac{1}{\sqrt{2}}(\left\vert f_{c}\right\rangle \left\vert
0\right\rangle -i\left\vert g_{c}\right\rangle \left\vert
1\right\rangle )\text{.}
\end{equation}%
Leaving the first cavity, the control atom is sequentially subjected
to two classical pulse in the Ramsey zone $R_{1}$. The first one is
tuned to induce the transition $\left\vert g_{c}\right\rangle
\longrightarrow i\left\vert
e_{c}\right\rangle $ and the second is for inducing the transformation $%
\left\vert e_{c}\right\rangle \longrightarrow \frac{1}{\sqrt{2}}(\left\vert
f_{c}\right\rangle -\left\vert e_{c}\right\rangle )$ and $\left\vert
f_{c}\right\rangle \longrightarrow \frac{1}{\sqrt{2}}(\left\vert
f_{c}\right\rangle +\left\vert e_{c}\right\rangle )$. The state (7) evolves
into
\begin{equation}
\frac{1}{2}(\left\vert f_{c}\right\rangle (\left\vert 0\right\rangle
+\left\vert 1\right\rangle )+\left\vert e_{c}\right\rangle (\left\vert
0\right\rangle -\left\vert 1\right\rangle ))\text{,}
\end{equation}%
which can be rewritten as follows:%
\begin{equation}
\frac{1}{2}(\left\vert f_{c}\right\rangle +\left\vert e_{c}\right\rangle
\sigma )(\left\vert 0\right\rangle +\left\vert 1\right\rangle )\text{,}
\end{equation}%
where $\sigma =\left\vert 0\right\rangle \left\langle 0\right\vert
-\left\vert 1\right\rangle \left\langle 1\right\vert $ is acting on the
collective mode of the atomic sample when the control atom is in the state $%
\left\vert e_{c}\right\rangle $.

Then the control atom is fed into the second cavity including another atomic
sample with the same level structure and atomic number. The state of the
second cavity is still in the vacuum and the collective mode of the second
atomic sample is also in the state $\left\vert 0\right\rangle $.\ Again, the
dispersive atom-cavity interaction leads to the oscillatory exchange of an
excitation between the control atom and the collective mode of the second
atomic sample. With the interaction time $t=\frac{\pi }{2\sqrt{N}\lambda _{c}%
}$ in the second cavity, the whole state of the atom system becomes
\begin{equation}
\frac{1}{2}(\left\vert f_{c}\right\rangle \left\vert 0\right\rangle
_{2}-i\left\vert g_{c}\right\rangle \left\vert 1\right\rangle
_{2}\sigma _{1})(\left\vert 0\right\rangle _{1}+\left\vert
1\right\rangle _{1})\text{.}
\end{equation}%
Here the subscripts $1,2$ are introduced to distinguish the atomic sample in
different cavities.

After leaving the second cavity, the control atom passes through the
second Ramsey zone $R_{2}$, where the control atom suffers in turns
the transition $\left\vert g_{c}\right\rangle \longrightarrow
i\left\vert e_{c}\right\rangle $ and the transformation $\left\vert
e_{c}\right\rangle \longrightarrow \frac{1}{\sqrt{2}}(\left\vert
f_{c}\right\rangle -\left\vert e_{c}\right\rangle )$ and $\left\vert
f_{c}\right\rangle \longrightarrow \frac{1}{\sqrt{2}}(\left\vert
f_{c}\right\rangle +\left\vert
e_{c}\right\rangle )$ by two classical pulses. Thus the state (10) becomes%
\begin{equation}
\frac{1}{2\sqrt{2}}(\left\vert f_{c}\right\rangle +\left\vert
e_{c}\right\rangle \sigma _{2})(\left\vert 0\right\rangle _{2}+\left\vert
1\right\rangle _{2}\sigma _{1})(\left\vert 0\right\rangle _{1}+\left\vert
1\right\rangle _{1})\text{.}
\end{equation}%
Continually, The control atom passes in sequence through the cavities
containing the same atomic sample and the Ramsey zones $C_{3}$, $R_{3}$, $%
C_{4}$, $R_{4}$, ..., $C_{K-1}$, $R_{K-1}$ as illustrated in the Fig. 1. The
whole state of the atom system evolves into
\begin{equation}
\frac{1}{\sqrt{2^{K}}}(\left\vert f_{c}\right\rangle +\left\vert
e_{c}\right\rangle \sigma _{K-1})(\left\vert 0\right\rangle
_{K-1}+\left\vert 1\right\rangle _{K-1}\sigma _{K-2})...(\left\vert
0\right\rangle _{2}+\left\vert 1\right\rangle _{2}\sigma _{1})(\left\vert
0\right\rangle _{1}+\left\vert 1\right\rangle _{1})\text{.}
\end{equation}%
Then the control atom is subjected to an extra Ramsey zone $R_{E}$
before entering the $K$th cavity, where a transition $\left\vert
f\right\rangle \leftrightarrow -i\left\vert g\right\rangle $ is
induced by one classical
pulse. The state becomes%
\begin{equation}
\frac{1}{\sqrt{2^{K}}}(-i\left\vert g_{c}\right\rangle +\left\vert
e_{c}\right\rangle \sigma _{K-1})(\left\vert 0\right\rangle
_{K-1}+\left\vert 1\right\rangle _{K-1}\sigma _{K-2})...(\left\vert
0\right\rangle _{2}+\left\vert 1\right\rangle _{2}\sigma
_{1})(\left\vert 0\right\rangle _{1}+\left\vert 1\right\rangle
_{1})\text{.}
\end{equation}%
Finally, after the control atom passes through the $K$th cavity holding the $%
K$th atomic sample with the duration $t=\frac{\pi }{2\sqrt{N}\lambda _{c}}$,
the quantum state of the atom system becomes%
\begin{equation}
\frac{1}{\sqrt{2^{K}}}\left\vert g_{c}\right\rangle (\left\vert
0\right\rangle _{K}+\left\vert 1\right\rangle _{K}\sigma _{K-1})(\left\vert
0\right\rangle _{K-1}+\left\vert 1\right\rangle _{K-1}\sigma
_{K-2})...(\left\vert 0\right\rangle _{2}+\left\vert 1\right\rangle
_{2}\sigma _{1})(\left\vert 0\right\rangle _{1}+\left\vert 1\right\rangle
_{1})\text{,}
\end{equation}%
where the control atom is disentangled with the atomic samples. Neglecting
the control atom state, we get a one-dimensional cluster state consisting of
$K$ atomic samples by encoding the vacuum state and one-excitation state of
the collective mode of the atomic sample as the logic zero and one of qubit.

\begin{figure}[tbp]
\includegraphics[width=8cm,height=2.5cm]{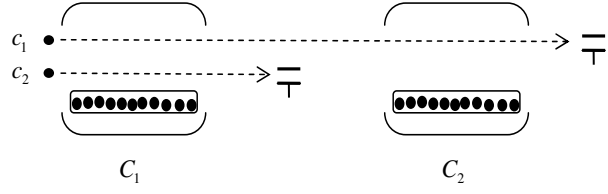}
\caption{Set-up for generating a two-dimensional cluster state. Firstly the
control atom $c1$ is sent through the two cavities $C_{1}$ and $C_{2}$ and
detected, then the control atom $c2$ is detected after passing through the
cavity $C_{1}$. }
\end{figure}

Up to now, one-dimensional cluster state of the atomic sample is generated,
yet it is insufficient for a practical task of computation on the one-way
quantum computer. We have to create various cluster state with different
shapes. Next, we describe a way to combine two one-dimensional cluster
states into a new two-dimensional cluster state. The schematic setup for
connecting two one-dimensional cluster states is given in Fig. 3, where the
atomic sample $i$ $(i=1,2)$ in the cavity $i$ is a node of a cluster state $%
\left\vert \Phi _{i}\right\rangle $ of atomic ensembles.
\begin{equation}
\left\vert \Phi _{i}\right\rangle =\frac{1}{\sqrt{2}}(\left\vert
0\right\rangle _{i}\left\vert \Psi _{i1}\right\rangle +\left\vert
1\right\rangle _{i}\left\vert \Psi _{i2}\right\rangle )\text{,}
\end{equation}%
which means the $i$th atomic sample is entangled with other atomic
samples, and the $\left\vert \Psi _{i1}\right\rangle $ and
$\left\vert \Psi _{i2}\right\rangle $ represent arbitrary normalized
states of the others. To connect two cluster states, a control atom
$c1$ is prepared in the quantum state $\frac{\left\vert
f_{c1}\right\rangle +\left\vert g_{c1}\right\rangle }{\sqrt{2}},$
and then is sent sequentially through two
cavities $1$ and $2$ with the same interaction $t=\frac{\pi }{\sqrt{N}%
\lambda _{c}}$. After the control atom $c1$ exits from the cavity $2$,
following the evolution described in the Eq.[6] the state of the atom system
becomes%
\begin{eqnarray}
&&\frac{1}{2\sqrt{2}}[\left( \left\vert f_{c1}\right\rangle +\left\vert
g_{c1}\right\rangle \right) \otimes \left( \left\vert 0\right\rangle
_{1}\left\vert 0\right\rangle _{2}\left\vert \Psi _{11}\right\rangle
\left\vert \Psi _{21}\right\rangle +\left\vert 1\right\rangle _{1}\left\vert
1\right\rangle _{2}\left\vert \Psi _{12}\right\rangle \left\vert \Psi
_{22}\right\rangle \right) \\
&&+\left( \left\vert f_{c1}\right\rangle -\left\vert g_{c1}\right\rangle
\right) \otimes \left( \left\vert 0\right\rangle _{1}\left\vert
1\right\rangle _{2}\left\vert \Psi _{11}\right\rangle \left\vert \Psi
_{22}\right\rangle +\left\vert 1\right\rangle _{1}\left\vert 0\right\rangle
_{2}\left\vert \Psi _{12}\right\rangle \left\vert \Psi _{21}\right\rangle
\right) ]\text{.}  \nonumber
\end{eqnarray}%
Then the state of the control atom $c1$ is detected by passing through the
atom $c1$ through the classical microwave field zone and field ionization
counters. When the state $\frac{\left\vert f_{c1}\right\rangle +\left\vert
g_{c1}\right\rangle }{\sqrt{2}}$ of the atom $c1$ is confirmed, the state of
two atomic-sample clusters is projected into%
\begin{equation}
\frac{1}{\sqrt{2}}\left( \left\vert 0\right\rangle _{1}\left\vert
0\right\rangle _{2}\left\vert \Psi _{11}\right\rangle \left\vert \Psi
_{21}\right\rangle +\left\vert 1\right\rangle _{1}\left\vert 1\right\rangle
_{2}\left\vert \Psi _{12}\right\rangle \left\vert \Psi _{22}\right\rangle
\right) \text{.}
\end{equation}%
Otherwise, the connection is failed.

The second control atom $c2$ prepared in the state $\left\vert
g_{c2}\right\rangle $ is fed into the cavity $1$ in the following step.
After the duration time $t=\frac{\pi }{2\sqrt{N}\lambda _{c}}$, the state of
the atom system becomes%
\begin{equation}
\frac{1}{\sqrt{2}}\left\vert 0\right\rangle _{1}\left( \left\vert
g_{c2}\right\rangle \left\vert 0\right\rangle _{2}\left\vert \Psi
_{11}\right\rangle \left\vert \Psi _{21}\right\rangle -i\left\vert
e_{c2}\right\rangle \left\vert 1\right\rangle _{2}\left\vert \Psi
_{12}\right\rangle \left\vert \Psi _{22}\right\rangle \right)
\text{,}
\end{equation}%
where the atomic sample $1$ is extricated from entanglement with others
samples. When the control atom $c2$ flies out from the cavity $1$, one
detects its state in the basic of $\{\frac{\left\vert g_{c2}\right\rangle
\pm i\left\vert e_{c2}\right\rangle }{\sqrt{2}}\}$. If the state $\frac{%
\left\vert g_{c2}\right\rangle -i\left\vert e_{c2}\right\rangle
}{\sqrt{2}}$ is announced, the above state (18) is projected into
\begin{equation}
\frac{1}{\sqrt{2}}\left( \left\vert 0\right\rangle _{2}\left\vert \Psi
_{11}\right\rangle \left\vert \Psi _{21}\right\rangle +\left\vert
1\right\rangle _{2}\left\vert \Psi _{12}\right\rangle \left\vert \Psi
_{22}\right\rangle \right) \text{,}
\end{equation}%
which is the standard form of a two-dimensional cluster state with
the atomic sample $2$ being the node. While the state
$\frac{\left\vert g_{c2}\right\rangle +i\left\vert
e_{c2}\right\rangle }{\sqrt{2}}$ is announced, the state (18)
collapses into
\begin{equation}
\frac{1}{\sqrt{2}}\left( \left\vert 0\right\rangle _{2}\left\vert \Psi
_{11}\right\rangle \left\vert \Psi _{21}\right\rangle -\left\vert
1\right\rangle _{2}\left\vert \Psi _{12}\right\rangle \left\vert \Psi
_{22}\right\rangle \right) \text{,}
\end{equation}%
which can be transformed into the form in Eq.(19) by local operation.
Therefore, we connect two one-dimensional cluster state of the atomic
samples into a two-dimensional one with the success probability $1/2$ and
one-qubit loss in the cluster.

In summary, we have taken the idea \cite{zheng} that the dynamics of
a control atom and an atomic sample interacting dispersively with a
cavity field can be described by the Jaynes-Cummings model and the
collective mode of the atomic sample can be analogous with a bosonic
mode. By analogizing the behavior of the atomic sample with the one
of the cavity field, we have proposed a scheme to generate the
cluster state of the atomic samples in the way similar to that in
Ref.\cite{zou}. In recent years the manipulation on one Rydberg atom
and clouds of Rydberg atoms has been demonstrated in the
laboratory by several groups in many remarkable experiments \cite%
{Zoller,Zoller2,JM,5}. We can choose the three circular levels with
principal quantum numbers 51, 50 and 49 to embody the states $\left\vert
e\right\rangle $, $\left\vert g\right\rangle $, $\left\vert f\right\rangle $%
, respectively. The $\left\vert g\right\rangle \leftrightarrow
\left\vert e\right\rangle $ and $\left\vert g\right\rangle
\leftrightarrow \left\vert f\right\rangle $ transitions are at 51.1
and 54.3 GHz, respectively. The radiative lifetimes of $\left\vert
e\right\rangle $, $\left\vert g\right\rangle $, $\left\vert
f\right\rangle $ is of the order of 30 ms. The coupling constant of
the atoms to the cavity is 25 kHz. As analyzed in Ref.\cite{zou},
the radiative time is enough for the atom-cavity interaction and
atom crossing the classical field. And during the whole process the
cavities in the proposed scheme remain in the vacuum, the cavity
loss can be negligible. Based on the present cavity QED techniques,
the proposed scheme to generate cluster state of the atomic
ensembles will be realizable in the future.

This work was supported by the National Natural Science Foundation of China
under Grant. Nos. 10404007 and 60578055, and the State Key Development
Program for Basic Research of China (Grant No. 2007CB925204 and
2009CB929604).

\end{document}